\documentclass[12pt,preprint]{aastex}
\usepackage{graphicx}
\usepackage{amssymb}

\makeatletter

\providecommand{\tabularnewline}{\\}

\usepackage{times}
\usepackage{amsmath}

\AtBeginDocument{
  
}

\makeatother

\begin{document}
\global\long\def\Mbh{M_{\bullet}}
\global\long\def\ms{m_{\star}}
\global\long\def\Ns{N_{\star}}
\global\long\def\Ms{M_{\star}}
\global\long\def\Rs{R_{\star}}
 \global\long\def\tvRR{T_{\mathrm{vRR}}}
\global\long\def\tsRR{T_{\mathrm{sRR}}}
\global\long\def\xRR{x_{RR}}
\global\long\def\Rlso{r_{\mathrm{lso}}}
\global\long\def\Plso{P_{\mathrm{lso}}}
\global\long\def\Nlso{N_{\mathrm{lso}}}
\global\long\def\Mo{M_{\odot}}
\global\long\def\Mdot{\dot{m}}

%\slugcomment{Draft version 8.2 \today}

\title{Accretion disk warping by resonant relaxation: the case of maser
disk NGC\,4258}

\shorttitle{Maser disk warping by resonant relaxation}

\author{Michal Bregman and Tal Alexander}

\shortauthors{Bregman \& Alexander}

\affil{Faculty of Physics, Weizmann Institute of Science, P.O. Box 26, Rehovot
76100, Israel}
\begin{abstract}
The maser disk around the massive black hole (MBH) in active galaxy
NGC\,4258 exhibits an $O(10^{\circ})$ warp on the $\mathrm{O(}0.1\,\mathrm{pc)}$
scale. The physics driving the warp are still debated. Suggested mechanisms
include torquing by relativistic frame dragging or by radiation pressure.
We propose here a new warping mechanism: resonant torquing of the
disk by stars in the dense cusp around the MBH. We show that resonant
torquing can induce such a warp over a wide range of observed and
deduced physical parameters of the maser disk.
\end{abstract}


\section{Introduction}

There is compelling evidence that massive black holes (MBHs) of mass
$10^{6}\, M_{\odot}\!\lesssim\!\Mbh\!\lesssim\!\mathrm{few}\!\times\!10^{9}\, M_{\odot}$
exist in the centers of most galaxies \citep{fer+00,geb+03,shi+03}.
The MBHs acquired most of their mass by efficient luminous accretion
\citep{sol82,yuq+02}, which implies accretion from an optically thick,
geometrically thin accretion disk \citep{sha+73}. On larger spatial
scales, such disks may fragment and form stars on disk-like orbits
\citep{kol+80,shl+89,goo03}, as observed in the Galactic Center \citep{bel+04,pau+06}
and elsewhere \citep{piz+02}. Gas disks, often marginally stable
and with properties similar to star disk progenitors \citep{mil+04},
are detected in the radio by $\mathrm{H_{2}O}$ maser emission (for
a list of known circum-nuclear maser disks, see \citealt{mal02};
for more candidates, see \citealt{bra+08}). 

The interest in processes that affect accretion disks stems from the
possible implications for MBH growth, for the luminosity evolution
of Active Galactic Nuclei (AGN) and for nuclear star formation. Conversely,
the disk properties may provide information about dynamical processes
that operate very close to the MBH, and may be relevant for other
phenomena there, such as tidal interactions with the MBH or the emission
of gravitational waves \citep[e.g.][]{ale05,ale07}. 

Maser-emitting nuclear accretion disks are unique, clean probes of
the near environment of MBHs. The first discovered and best studied
maser disk, NGC\,4258, is also the thinnest and most Keplerian (to
better than $1\%$, \citealt{mal02}) nuclear disk yet discovered.
Radio $\mathrm{H_{2}O}$ maser observations of the disk morphology,
velocities and accelerations allow accurate measurements of the MBH
mass and the distance to the host galaxy \citep{her+96}. Together
with optical and X-ray observations, these provide estimates the mass
accretion rate through the disk \citep{neu+95}. The NGC\,4258 maser
disk shows a clear $O(10^{\circ})$ warp on the $\mathrm{few\times}0.1\,\mathrm{pc}$
scale \citep{her+96}, whose origin is still a matter of debate. Warps
are possibly also observed in the maser disks of Circinus \citep{gre+03}
and NGC 3393 \citep{kon+08}.

Several mechanisms were proposed to explain the warped disk in NGC~4258.
The \citet{bar75} (BP) effect (torquing of a viscous disk by the
General Relativistic frame-dragging) can warp a disk that is initially
misaligned with the MBH spin by dragging its inner regions to the
MBH equatorial plane \citep{lod+06,cap+07,mar08}. Instability of
a thin disk to radiation pressure from the central source can result
in the amplification of a small pre-existing warp \citep{pri96,mal+96}.
The disk's self-gravity can support equilibrium warped disk configurations
\citep{ulu+07}, if these are excited by other means. One feature
shared by these mechanisms is that they require some initial asymmetry
in the disk or disk/MBH alignment for the warp to develop.

Here we consider the implications of the dense stellar cusp around
the gas disk. Various dynamical scenarios predict the formation of
steep stellar density cusps $n(r)\!\propto\!(r/r_{h}){}^{-\gamma}$
($0.5\lesssim\!\gamma\!\lesssim2.5$) within the MBH radius of influence
$r_{h}$ \citep[e.g.][]{bah+77}. The empirical $\Mbh/\sigma$ correlation
implies that the cusp density scales as $\Mbh^{-1/2}$, so that the
relatively low-mass MBHs, where maser disks are found, are surrounded
by a very dense stellar cusp \citep{ale07}. Strong mass segregation
(\citealt{ale+09}) further concentrates stellar-mass black holes
(SBHs) to the center. The stars in the cusp move in a nearly spherical
potential dominated by the MBH, and therefore orbit on nearly fixed
planes. This symmetry leads to the rapid torquing of any test particle
by the process of resonant relaxation (RR) (\citealt{rau+96} (RT96);
\citealt{hop+06a}). In particular, gas streams on circular orbits
undergo {}``vector RR'', which changes the orbital plane, but not
their eccentricity. RR thus induces warps by exchanging angular momentum
between the stars and the disk.

This \emph{letter} is organized as follows. RR dynamics are summarized
in \S \ref{s:RRdyn}. An analytical model for RR torquing of a disk
is described in \S \ref{s:RRdisk}, and applied to the NGC\,4258
maser in \S  \ref{s:model}. Our results are presented in \S \ref{s:results}
and summarized in \S \ref{s:discuss}.

\section{Resonant relaxation dynamics}

\label{s:RRdyn}

RR is a rapid relaxation mechanism of the angular momentum $\mathbf{L}$,
which operates in near-symmetric potentials that suppress the evolution
of the stellar orbits (e.g. fixed ellipses in a Kepler potential,
or planar rosettes in a spherical potential). In such systems, the
residual torque on a test mass by the orbit-averaged mass distributions
of $\Ns$ stars of mass $\Ms$ within distance $r$ from the center,
$|\mathbf{T}|\!\sim\!\Ns^{1/2}G\Ms/r$, remains constant on timescales
shorter than the coherence time $t_{0}$, as long as perturbations
due to deviations from the perfect symmetry remain small. The change
in $\mathbf{L}$ then grows coherently as $\propto\! t$ on timescales
$P\!\ll\! t\!\ll\! t_{0}$, where $P\!=\!2\pi\sqrt{r^{3}/G(\Mbh+\Ns\Ms)}$
is the circular orbital period. When $t\!\gg\! t_{0}$, the large
accumulated change over the coherence time, $(\Delta L)_{0}\!=\!|\mathbf{T}|t_{0}$,
becomes the step size of a rapid random walk, $(\Delta L/L_{c})(t)\!=\!\left[(\Delta L)_{0}/L_{c}\right]\sqrt{t/t_{0}}\!\equiv\!\sqrt{t/T_{\mathrm{RR}}}$,
where $L_{c}\!=\!\sqrt{G(\Mbh+\Ns\Ms)r}$ is the circular specific
angular momentum for the test mass' orbital energy, and $T_{\mathrm{RR}}$
is thereby defined as the RR timescale, when $|\Delta L|/L_{c}\!=\!1$.
The slower the loss of coherence ({}``quenching''), the more efficient
is RR (shorter $T_{\mathrm{RR}}$). RR can be orders of magnitude
faster than non-coherent 2-body relaxation.

In a near-Keplerian potential the orbits are nearly fixed ellipses,
and the coherence time is set by the faster of General Relativistic
(GR) precession or precession by the potential of the enclosed stellar
mass. In this case RR can change both the direction and magnitude
of $\mathbf{L}$ ({}``scalar RR''). However, the torques along $\mathbf{L}$
fall to zero for a test mass on a circular orbit \citep{gur+07},
and so circular orbits, such as those of gas streams in an accretion
disk, remain circular ($\Delta\mathbf{T}_{\parallel}\!=\!0$), but
their orientation evolves rapidly ($\Delta\mathbf{T}_{\perp}\!\neq\!0$,
{}``vector RR'') until RR itself randomizes the orbital planes to
a degree where coherence is lost ({}``self-quenching''). Vector
RR is not quenched by precession ($\Delta\mathbf{T}_{\perp}\!\neq\!0$
between orbit-averaged rosettes), and so it can operate both very
near the MBH, where GR precession is fast, and far from the MBH where
the potential is spherical but no longer Keplerian. The coherence
time for the self-quenched vector RR is $t_{0}\!=\! A_{0}L_{c}/|\mathbf{T}|\!=\! A_{0}\sqrt{N_{\star}}P/\mu$,
where $\mu\!=\!\Ns\Ms/(\Mbh+\Ns\Ms)$ and where $A_{0}$ is an order
unity factor. On timescales $t\!\ll\! t_{0}$, 

\begin{equation}
\frac{|\Delta\mathbf{L}|}{L_{c}}=\frac{\beta_{v}}{\sqrt{\Ns}\mu}\frac{t}{P}\simeq\beta_{v}\frac{\Ms}{\Mbh}\sqrt{\Ns}\frac{t}{P}\,\,\,\,\,\,\,(\Ns\Ms\!\ll\!\Mbh)\,.\label{e:tvRRlin}\end{equation}
where $\beta_{v}$ is an order-unity factor (recent $N$-body simulations
indicate that $\beta_{v}\!\simeq\!1.8$, \citealt{eil+09}). The vector
RR timescale is then

\begin{equation}
\tvRR=\frac{1}{\beta_{v}^{2}A_{0}}\frac{\Ns^{1/2}}{\mu}P\simeq\frac{1}{\beta_{v}^{2}A_{0}}\left(\frac{\Mbh}{\Ms}\right)\frac{P}{\sqrt{\Ns}}\,\,\,\,(\Ns\Ms\!\ll\!\Mbh)\,.\label{e:tvRR}\end{equation}
For the purpose of numeric evaluation, we assume below $\beta_{v}^{2}A_{0}\!=\!2$.
Note that for vector RR, $T_{\mathrm{vRR}}\!\sim\! t_{0}$. 

A change of $w\!\equiv\!|\Delta\mathbf{L}|/L_{c}$ in the angular
momentum of a gas ring in an accretion disk ({}``warp factor'' $w$),
expressed by the change to the ring normal $\Delta\mathbf{\hat{n}}\!=\!(\theta,\varphi)$,
corresponds to a warp inclination angle $\theta$ via $w^{2}=2(1-\cos\theta)$,
For $w\ll1$, $w\simeq\theta$ rad. By definition, $\theta\!=\!60^{\circ}$
($w\!=\!1$) over the vector RR timescale $T_{\mathrm{vRR}}$. RR
warping by a factor $w\!\ll\!1$ occurs on a timescale shorter than
the coherence time ($w\!\propto\! t$), and so the time required to
develop such a small local warp inclination is $t(w)\!\sim\! wt_{0}\!\sim\! wT_{\mathrm{vRR}}$. 

The residual RR torque $\mathbf{T}(r)$ changes gradually with $r$,
as the uncorrelated stars that effectively contribute to it vary.
We estimate the RR spatial coherence factor $f_{c}\!=\! r_{2}/r_{1}$
as the distance ratio over which the enclosed stellar numbers satisfy
$2\sqrt{N(r_{1})}\!=\!\sqrt{N(r_{2})}$, which translates to $f_{c}\!=\!2^{2/(3-\gamma)}$.
The observed NGC\,4258 maser disk extends over $R_{2}/R_{1}\!=\!2$,
and the assumed $\gamma\!=\!1.5$ stellar cusp there corresponds to
$f_{c}\!\simeq\!2.5$ (\S \ref{s:model}), which implies that the
warping torques, and thus $\Delta\mathbf{\hat{n}}_{1}$ and $\Delta\mathbf{\hat{n}}_{2}$,
are approximately uncorrelated (section \ref{s:model}). The total
observed warp angle $\omega\!=\!\cos^{-1}(\Delta\hat{\mathbf{n}}_{1}\cdot\Delta\hat{\mathbf{n}}_{2})$
is due to the difference between the local warp angles at the disk's
inner and outer edges. The relative magnitude of the warp factors
is related by Eq. (\ref{e:tvRRlin}), $f_{w}\!=\! w_{2}/w_{1}\!=\!(R_{2}/R_{1})^{-\gamma/2}$
($f_{w}=0.6$ for $\gamma=1.5$). In the limit of small warp inclination
angles, as is the case here, $f_{w}\!\simeq\!\theta_{2}/\theta_{1}$
and $\omega\!\simeq\!\theta_{1}\sqrt{1-2f_{w}\cos(\varphi_{1}-\varphi_{2})+f_{w}^{2}}$.
The maser disk in NGC~4258 displays a $\omega=8^{\circ}$ relative
warp between the limits of the masing region at $R_{1}=0.14\,\mathrm{pc}$,
$R_{2}=0.28\,\mathrm{pc}$ \citep{her+96}, which after averaging
over uncorrelated $\varphi_{1}$ and $\varphi_{2}$, implies that
locally, $\theta_{1}\simeq10^{\circ}$ and $\theta_{2}\simeq6^{\circ}$.

\section{Accretion disk warping by resonant relaxation}

\begin{figure*}[t]
\begin{tabular}{cc}
$\negthickspace\negthickspace\negthickspace\negthickspace\negthickspace\negthickspace\negthickspace\negthickspace\negthickspace\negthickspace\negthickspace\negthickspace$\includegraphics[width=0.557\linewidth]{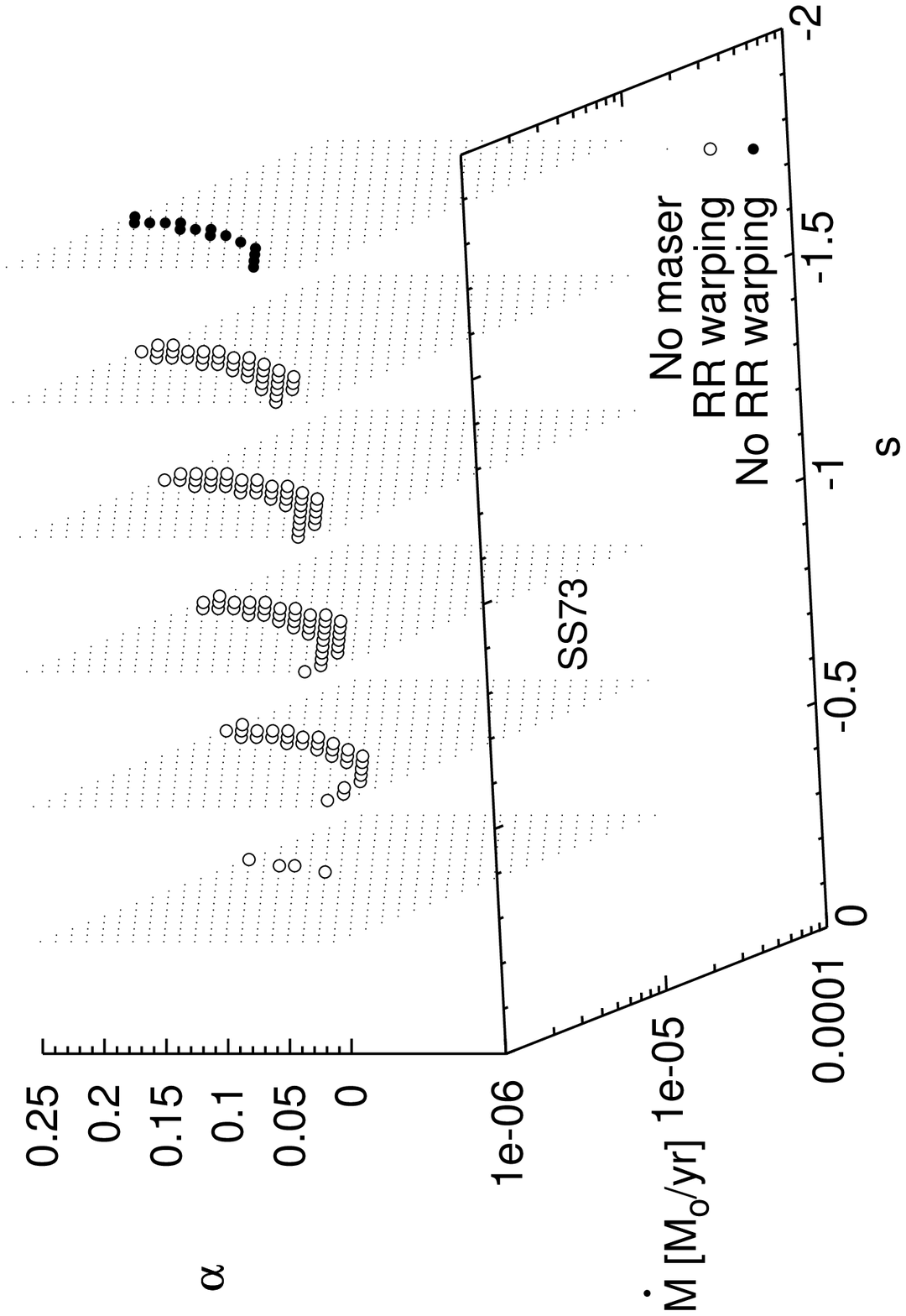} & $\negthickspace\negthickspace\negthickspace\negthickspace\negthickspace\negthickspace\negthickspace\negthickspace\negthickspace\negthickspace\negthickspace\negthickspace$\includegraphics[width=0.5\linewidth]{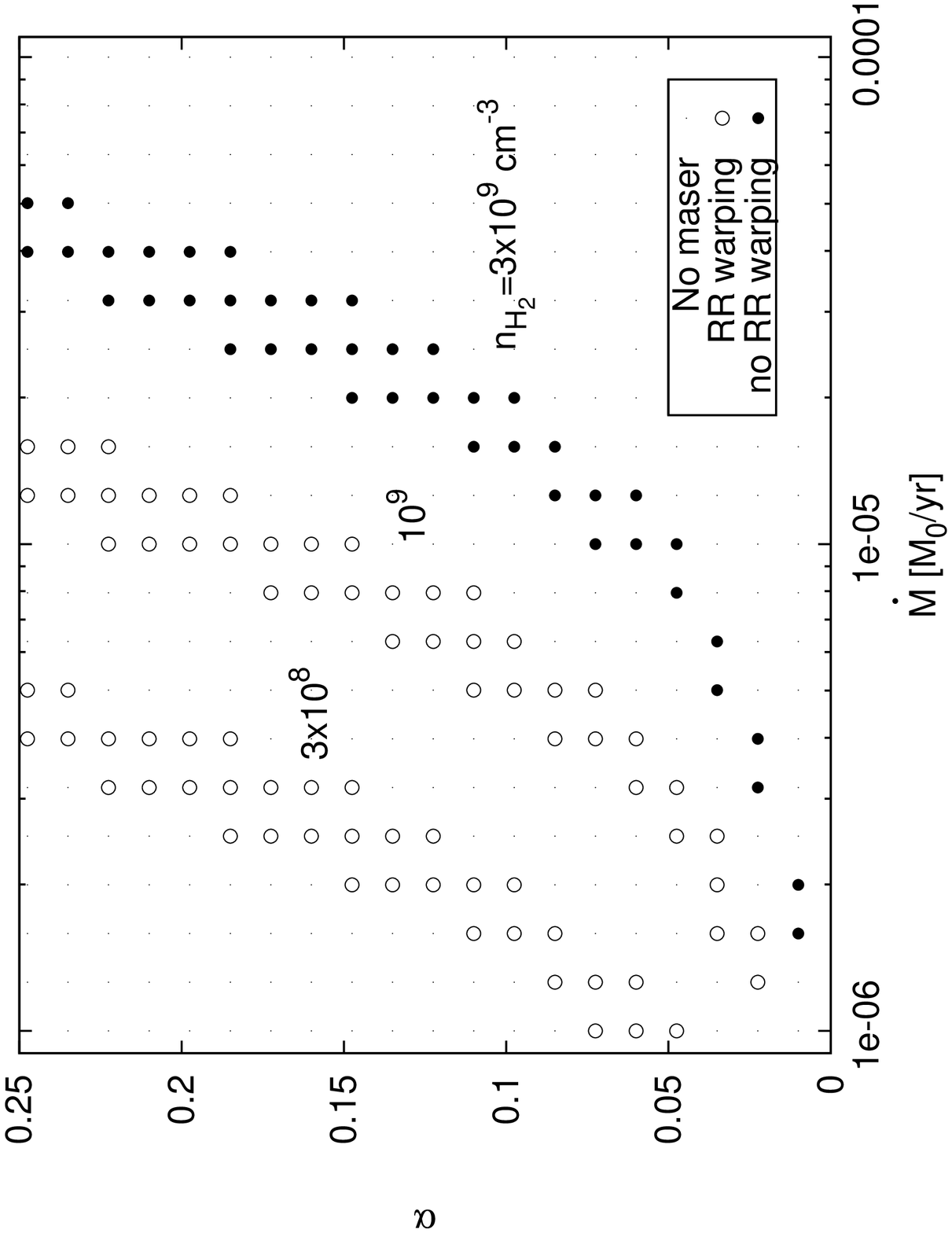}\tabularnewline
\end{tabular}

\caption{\label{f:warp}RR disk warping in NGC~4258 disk parameter space ($\alpha$,
$\dot{M}$, $s$) . Left: Range of surface density slopes $s$ with
$n_{\mathrm{H}_{2}}(R_{1})\!=\!10^{9}\,\mathrm{cm^{-3}}$. Parameter
space regions where RR warping by $\omega\!\ge\!8^{\circ}$ between
$ $$0.14\,\mathrm{pc}$ and $0.28\,\mathrm{pc}$ can be achieved
as observed, are marked by open circles, regions where RR is not efficient
enough are marked by filled circles, and regions where the physical
conditions in the disk are incompatible with $\mathrm{H_{2}O}$ maser
production (\S \ref{ss:diskmodel}) are marked by dots. Right: The
same, for $s\!=\!-0.75$ (SS73 $\alpha$-disk model) and for $n_{\mathrm{H}_{2}}(R_{1})\!=\!3\times10^{8},$
$10^{9}$ and $3\times10^{9}\,\mathrm{cm^{-3}}$.}

\end{figure*}

\label{s:RRdisk}

The gas in a thin accretion disk flows slowly into the MBH on nearly
circular orbits. Over time, the orbit associated with a ring element
of the disk shrinks, and the torques on it change. RR can significantly
affect the ring only if the typical RR timescale, $T_{\mathrm{vRR}}$
is shorter than the radial inflow timescale $t_{r}\!\sim\! R^{2}/\nu$,
which is set by the disk's tangential in-plane viscosity coefficient,
$\nu$, where $R$ is the radius along the disk's mid-plane. The presence
of a warp also induces a vertical shear viscosity $\nu_{2}$, which
resists the external torques by diffusing the warp on a timescale
$t_{w}=R^{2}/\nu_{2}$ \citep{pri92,lod+07}. Effective warping thus
also requires that $T_{\mathrm{vRR}}$ be shorter than the warp diffusion
timescale. In addition to these timescale constraints, the torquing
stars must carry enough residual angular momentum to exchange with
the disk and warp it. Therefore, both a timescale condition (\S \ref{ss:RRT})
and an angular momentum condition (\S \ref{ss:RRL}) must be satisfied
for RR torquing to occur.

\subsection{RR timescale condition}

\label{ss:RRT}

The vector RR timescale (Eq. \ref{e:tvRR}) estimated above for stars
enclosed within radius $r$ and acting on a mass orbiting at the same
typical radius $r$, can be easily generalized to a mass (disk ring
element) at a different radius $R$, by noting that $T_{\mathrm{vRR}}$
scales with the torque as $\propto|\mathbf{T}|^{-2}$, and that the
typical torque on the ring element is decreased by a factor $r/R$
when $R\!>\! r$ and by a factor $R/r$ when $R\!<\! r$, 

\begin{equation}
\tvRR(r,R)=\tvRR(r)(r/R)^{2\Theta}\propto r^{2\Theta+\gamma/2}\,,\label{e:vRRg}\end{equation}
where $\Theta\!=\!\mathrm{sgn}(r-R)$, and the last proportionality
holds in the Keplerian limit. In order to efficiently warp the disk
by a factor $w$, $\tvRR$ must be shorter by at least some factor
$\epsilon_{t}\!\le\!1$ than the faster of the two viscous processes,
$t_{rw}=\min(t_{r},t_{w})$,

\begin{equation}
\epsilon_{t}t_{rw}(R)\ge w\tvRR(r,R)\,.\label{e:tcond}\end{equation}
An equality in Eq. (\ref{e:tcond}) defines \emph{an upper limit}
$r_{t}^{(+)}$ for a given $R\!<\! r$ and a \emph{lower limit} $r_{t}^{(-)}$
for $R\!>\! r$ (assuming $0\!<\!\gamma\!<\!4$; recall that $0.5\!<\!\gamma\!<\!2.5$
for realistic cusp models) 

\begin{equation}
\frac{r_{t}^{(\Theta)}}{\Rlso}=\left[\sqrt{\Nlso}\frac{\Ms}{\Mbh}\frac{(\epsilon_{t}/w)t_{rw}(R)}{\Plso}\left(\frac{R}{\Rlso}\right)^{2\Theta}\right]^{1/(2\Theta+\gamma/2)}\,,\end{equation}
where $N_{\mathrm{lso}}$ is the formally extrapolated number of stars
within the last stable circular orbit $r_{\mathrm{lso}}$, and $P_{\mathrm{lso}}$
is the formal Keplerian orbital period there. When there exists a
solution with $r_{t}^{(-)}\!<\! r_{t}^{(+)}$, then stars in the volume
$r_{t}^{(-)}\!\le\! r\!\le\! r_{t}^{(+)}$ induce fast enough vector
RR on the disk at $R$ to warp it by a factor $w$. 

Note that the timescale condition implies that the RR torque dominates
over the viscous torques, since $|\mathbf{T}_{\mathrm{RR}}|\sim L_{c}/T_{\mathrm{vRR}}\!>\! L_{c}/t_{rw}\!\sim\!\max(|\boldsymbol{\mathbf{T}}_{r}|,|\mathbf{T}_{w}|)$.

\subsection{RR angular momentum condition}

\label{ss:RRL}

Efficient torquing of the disk mass in a ring $(R,R+\Delta R)$ by
stars in a spherical shell $(r,r+\Delta r)$ requires that the disk's
angular momentum be smaller by at least some factor $\epsilon_{L}\!\le\!1$
than the random residual angular momentum carried by the stars, assumed
to be isotropically distributed (this a conservative criterion, since
it neglects the possibility of warping by transfer of angular momentum
from the disk to the stars). The total angular momentum in both the
disk and the stellar cusp is dominated by the large scales (e.g. a
disk with surface density $\Sigma\!\propto\! R^{s}$ (\S \ref{ss:diskmodel})
and temperature that falls with $R$ has $s\!>\!-3/2$ (Eq. \ref{e:HvRcs})
and $L_{d}\Delta M_{d}\!\propto\! R^{3/2+s}\Delta R$, where $M_{d}$
is the disk mass enclosed within $R$; a stellar cusp with $\gamma\!<\!2$
(\S \ref{ss:GNmodel}) has $L_{\star}\Delta(\sqrt{\Ns}\Ms)\!\propto\! r^{1-\gamma/2}\Delta r$).
We therefore approximate the angular momentum criterion by requiring
that the angular momentum in the disk within $R$ be smaller than
the residual angular momentum in the stars within $r$, 

\begin{equation}
wM_{d}(R)L_{c}(R)\le\epsilon_{L}\sqrt{\Ns(r)}\Ms L_{c}(r)\propto\left(r/\Rlso\right)^{2-\gamma/2}\,.\label{e:jcond}\end{equation}
 An equality in Eq. (\ref{e:jcond}) defines\emph{ a lower limit}
$r_{L}$ (assuming $\gamma<4$)\begin{equation}
\frac{r_{L}}{\Rlso}=\left[\frac{1}{(\epsilon_{L}/w)\sqrt{\Nlso}}\frac{M_{d}(R)}{\Ms}\sqrt{\frac{R}{\Rlso}}\right]^{1/(2-\gamma/2)}\,.\end{equation}
When there exists a solution such that $r^{(-)}\!=\!\max(r_{t}^{(-)},r_{L})\!<\! r_{t}^{(+)}$,
then the stars in the volume $r^{(-)}\!<\! r\!<\! r_{t}^{(+)}$ carry
enough residual angular momentum to significantly torque the disk
enclosed within $R$, on a short enough timescale. 

Note that the angular momentum condition implies that the RR torque
dominates over the disk's self-gravity, since the self-gravity torques
$|\mathbf{T}_{d}|\!\sim\! GM_{d}/r\!<\! G\sqrt{\Ns}\Ms/r\!\sim\!|\mathbf{T}_{\mathrm{RR}}|$.

\section{NGC\,4258 model}

\label{s:model}

In order to apply the RR torquing mechanism to the maser disk in NGC\,4258,
it is necessary to specify the stellar cusp density profile, the disk's
viscous timescales, $t_{rw}(R)$ (Eq. \ref{e:tcond}), and its enclosed
mass $M_{d}(R)$ (Eq. \ref{e:jcond}). We adopt here simple, observationally
motivated models of the cusp and disk.

\subsection{Galactic nucleus model}

\label{ss:GNmodel}

NGC\,4258 is a spiral galaxy at a distance of $7.2\pm0.3\,\mathrm{Mpc}$
with a $\Mbh\!\sim\!3.7\times10^{7}\, M_{\odot}$ central MBH \citep{her+99}.
The empirical $\Mbh/\sigma$ relation \citep{fer+00,geb+03,shi+03}
implies that $\sigma\!\simeq\!150\,\mathrm{km\, s^{-1}}$ and the
MBH's radius of dynamical influence is $r_{h}\!=G\Mbh/\sigma^{2}\!\simeq\!7$
pc, where the enclosed stellar mass is $\mu_{h}\Mbh$, with $\mu_{h}\!\sim\! O(1)$.
On the distances spanned by the maser spots, $R_{1}\!=\!0.14$ to
$R_{2}\!=\!0.28$ pc \citep{her+96}, main-sequence stars dominate
the population with a relatively flat $\gamma\!\sim\!1.5$ power-law
density profile, typical of the low-mass component in a mass-segregated
population \citep{bah+77,ale+09}. We model the central cusp of NGC\,4258
as a power-law cusp, $\Ns(<r)\!=\!\mu_{h}(\Mbh/\Ms)(r/r_{h})^{3-\gamma}$,
with $\gamma\!=\!1.5$, $r_{h}\!=\!7$ pc, $\mu_{h}\!=\!2$ (the formal
value for an MBH-less singular isothermal distribution) and $\Ms\!=\!1\,\Mo$
stars. The corresponding vector RR timescale is $T_{\mathrm{vRR}}\!\sim\!\mathrm{few\times}10^{7}$
yr across the disk.

\subsection{Maser disk model}

\label{ss:diskmodel}

Following \citet{cap+07}, we assume a power-law surface density profile
for the disk, $\Sigma(R)\!=\!\Sigma_{1}(R/R_{1})^{s}$. We assume
the structure equations of a stationary, geometrically thin, optically
thick Keplerian disk around a MBH of mass $\Mbh$ (e.g. \citealt{fra+02}).
These can be expressed in terms of the mass accretion rate $\dot{M}$,
the dimensionless viscosity parameter $\alpha\!\equiv\!\nu/c_{s}H$,
the surface density power-law index $s$ and $\rho(R_{1})$, the mid-plane
mass density of $\mathrm{H}_{2}$ at $R_{1}$ (the disk mass is assumed
to be mostly in $\mathrm{H}_{2}$). The normalization $\Sigma_{1}$
at $R_{1}$ is fixed by\begin{equation}
\Sigma^{3}=\dot{M}\rho^{2}\left/2\alpha\Omega\right.\,,\end{equation}
where $\Omega\!=\!\sqrt{G\Mbh/R^{3}}$ is the Keplerian frequency,
$H$ is the disk scale-height $H$, and $c_{s}$ is isothermal sound
speed $c_{s}$,

\begin{equation}
H^{2}\!=\!\dot{M}\!\left/\!2\pi\alpha\Sigma\Omega\right.,\,\,\,\,\, c_{s}^{2}\!=\!\dot{M}\Omega\!\left/\!2\pi\alpha\Sigma\right.\,.\label{e:HvRcs}\end{equation}
For an ideal gas $c_{s}\!=\!\sqrt{kT/\mu}$ with $T$ the gas temperature
and $\mu$ the mean molecular mass ($\mu\!=\!2m_{p}$ assumed). The
radial inflow and the warp diffusion viscous timescales are then \begin{equation}
t_{r}\!\sim\! R^{2}\left/\alpha c_{s}H\right.\,,\,\,\,\,\, t_{w}\!\sim\! R^{2}\left/\alpha_{2}c_{s}H\right.\,,\label{e:tv}\end{equation}
 where $\alpha_{2}\!=\!\max[f(\alpha),\alpha_{2\max}]$ with $f(\alpha)\!\simeq\!2(1+7\alpha^{2})/[\alpha(4+\alpha^{2})]$
\citep{ogi99} and $\alpha_{2\, max}\!\sim\!3$--$4$, with some uncertainty
\citep{lod+07}.

\citet{cap+07} combine observational and theoretical constraints
and deduce that the physical parameters of the NGC\,4258 maser disk
lie in the range $10^{-6}\!\lesssim\!\epsilon\dot{M}\!\lesssim\!10^{-4}\, M_{\odot}\,\mathrm{yr^{-1}}$,
where $\varepsilon$ is the radiative accretion efficiency (but see
higher estimate $\dot{M\!}\gtrsim\!10^{-3}\,\Mo\,\mathrm{yr^{-1}}$
for ADAF model, \citealt{las+96}), and $0.03\!\lesssim\!\alpha\!\lesssim\!0.2$.
They consider a range of possible disk profiles, $-2\!\leq\! s\!\leq\!0$.
 Efficient production of $\mathrm{H_{2}O}$ maser emission constrains
the disk's physical parameters: the $\mathrm{H_{2}}$ density, $10^{7}\,\mathrm{cm^{-3}}\!<\! n_{\mathrm{H}_{2}}\!<\!10^{11}\,\mathrm{cm^{-3}}$,
the gas temperature, $400\,\mathrm{K}<\! T\!<\!1000\,\mathrm{K}$,
and the gas pressure $10^{10}\,\mathrm{K\, cm^{-3}}\!<\! p/k\!<\!10^{13}\,\mathrm{K\, cm^{-3}}$
\citep{mal02}. In addition, the disk aspect ratio is constrained
by observations to $H/R\!\lesssim\!0.002$ \citep{mor08}. The disk
is gravitationally stable only if $M_{d}/\Mbh\!<\! H/R$.

\[
\]

\section{Results}

\label{s:results}

Figure (\ref{f:warp}) shows the region in the disk's $(\dot{M},\alpha,s,n_{H_{2}})$$ $
parameter space where masing is possible with the aspect ratio and
disk mass constraints (\S \ref{ss:diskmodel}), and where both the
timescale (Eq. \ref{e:tcond} with $\epsilon_{t}\!=\!1$, $\alpha_{2\max}\!=\!3$),
and angular momentum conditions (Eq. \ref{e:jcond} with $\epsilon_{L}\!=\!1$)
are satisfied for $\theta\!\ge\!10^{\circ}$ between $0.14$ and $0.28$
pc, so that RR torquing can warp the disk by at least $8^{\circ}$,
on average. We find that such an RR-induced warp is possible over
a wide range of the observed and deduced physical parameters of the
maser disk. RR warping is less efficient for disks with a very steep
density profile, high density or a high mass accretion rate. We conclude
that the RR mechanism can drive a moderate warp in the NGC\,4258
maser disk, but cannot substantially perturb or disrupt it. We also
find that the highest accretion rate consistent with masing is $\dot{M}\!<\!10^{-4}\,\Mo\,\mathrm{yr^{-1}}$,
which agrees with X-ray based estimates \citep{wat02}.

\section{Summary}

\label{s:discuss}

We have shown that the torques exerted by the Poisson fluctuations
in the distribution of the stars around the MBH in NGC\,4258 can
transfer momentum from the stars to the maser disk and excite the
warp observed in the disk on a timescale $wT_{\mathrm{vRR}}\sim O(10^{6}\,\mathrm{yr})$,
more efficiently than the BP effect ($t_{\mathrm{BP}}\!\gtrsim\!\mathrm{few}\times10^{9}\,\mathrm{yr}$,
\citealt{cap+07}) or radiation pressure (\citealt{gam+99}; see \citealt{cap+07}
and references therein). RR is inherent to the discreteness of the
stellar system, and thus does not depend on particular initial conditions
for the disk, such as the initial warp required for radiation pressure
warping, or the initial misalignment between the disk and the MBH
spin axis required by the BP effect. RR-induced warps are transient
and vary on the RR timescale $T_{\mathrm{vRR}}\!\sim\! O(10^{7}\,\mathrm{yr})$. 

The NGC\,4258 maser disk stands out in its well-defined morphology,
Keplerian dynamics and the available detailed high-quality observations.
These make it ideal for testing the RR torquing model. However, vector
RR operates also in non-Keplerian spherical potentials, and can warp
any disk configuration, as long as its mass is small enough and the
viscous timescales are long enough. It therefore likely that RR torquing
is relevant in other maser disk systems as well. 

Generally, RR torquing of accretion disks is also expected on much
smaller scales, driven by the strongly segregated cusp of stellar
mass black holes \citep{ale+09}. RR could thus potentially affect
the accretion rate and direction, and thereby MBH mass and spin evolution.

\acknowledgements{We thank S. Tremaine, J.-P. Lasota and the anonymous referee for
helpful discussions and comments. TA acknowledges support by ISF grant
928/06 and ERC Starting Grant 202996.}

\bibliographystyle{apj}
%\bibliography{maser}

\end{document}